\documentclass[a4paper,fleqn]{cas-sc}

\usepackage[numbers]{natbib}

\def\tsc#1{\csdef{#1}{\textsc{\lowercase{#1}}\xspace}}
\tsc{WGM}
\tsc{QE}
\tsc{EP}
\tsc{PMS}
\tsc{BEC}
\tsc{DE}

\newtheorem{thm}{Theorem}

\newdefinition{rmk}{Remark}
\newproof{pf}{Proof}

\begin{document}
\let\WriteBookmarks\relax
\def\floatpagepagefraction{1}
\def\textpagefraction{.001}
\shorttitle{Mathematical Validation of a Cancer Model}
\shortauthors{M.V. Reale et~al.}

\title [mode = title]{Mathematical Validation of a Cancer Model}                      



\author[1,3]{Marcela V. Reale}[orcid=0000-0002-9856-7501]
\ead{mreale@campus.ungs.edu.ar}

\credit{Conceptualization, Methodology, Formal analysis, Validation, Original draft preparation, Writing - Review \& Editing}

\author[2]{Gustavo Paccosi}[
                        orcid=0000-0002-0895-1957]

\credit{Conceptualization, Methodology, Formal analysis, Validation, Writing - Review \& Editing}

\author[1,3]{David H. Margarit}[
                        orcid=0000-0003-1946-0413]

\credit{Validation, Writing - Review \& Editing}

\author[1,3]{Lilia Romanelli}[
                        orcid=0000-0002-9272-6575]
\credit{Funding acquisition, Validation, Writing - Review \& Editing}

\affiliation[1]{organization={Instituto de Ciencias, Universidad Nacional de General Sarmiento}, addressline={J.M. Gutierrez 1150}, city={Los Polvorines}, postcode={B1613}, state={Buenos Aires}, country={Argentina}}

\affiliation[2]{organization={Instituto del Desarrollo Humano, Universidad Nacional de General Sarmiento}, addressline={J.M. Gutierrez 1150}, city={Los Polvorines}, postcode={B1613}, state={Buenos Aires}, country={Argentina}}

\affiliation[3]{organization={Consejo Nacional de Investigaciones Cient\'iificas y T\'ecnicas (CONICET)},addressline={Godoy Cruz 2290}, city={Ciudad Aut\'onoma de Buenos Aires}, postcode={C1425}, state={Buenos Aires}, country={Argentina}}

\cortext[cor1]{Corresponding author}

\begin{abstract}
Understanding cancer cell differentiation is essential for advancing its detection, diagnosis, and treatment. Mathematical models significantly contribute to this by providing a theoretical framework to understand the complex interactions between cancer stem cells, differentiated cancer cells, and immune system components. Such models depend on experimental data and computational simulations to predict tumor dynamics, offering insights into how different cell populations evolve over time. However, to ensure their realistic and consistent outcomes, rigorous mathematical analysis is required, including verification of solution uniqueness, stability, viability, positivity, and boundedness. Such validation guarantees that model's results can be used in both oncological research and clinical applications.
In this study, we conduct a comprehensive analysis of an integrative mathematical model of cancer cell differentiation, with a particular focus on its interactions with immune cells. The model captures the dynamic balance between cancer stem cell self-renewal, differentiation into mature tumor cells, and immune-mediated elimination. By employing analytical and numerical techniques, we assess the model's feasibility, stability, and long-term behavior under various biological conditions. Our findings demonstrate that immune system engagement can significantly influence tumor composition and growth, highlighting potential therapeutic targets. This work not only advances theoretical cancer modeling but also provides a foundation for future experimental validation and the development of combined differentiation-immunotherapy approaches. The results underscore the importance of interdisciplinary collaboration in the fight against cancer.
\end{abstract}

\begin{highlights}
\item Rigorous mathematical analysis of an integrative cancer cell differentiation model.
\item Validation of the model's feasibility, stability, positivity, and boundedness.
\item Interaction dynamics between cancer cells and immune system populations explored.
\item Foundational framework for advancing mathematical oncology research.
\end{highlights}


\maketitle

\section{Introduction}\label{sec1}

Cancer modelling has become indispensable in understanding and managing the disease, contributing significantly to its detection, diagnosis, and treatment across various stages. Unlike traditional approaches, mathematical models integrate experimental data from in vivo and in vitro studies, enhancing their applicability in contemporary and constantly evolving therapeutic contexts \cite{reviewmath}. Moreover, the predictive power of numerical simulations strengthens their clinical relevance by providing insights into cancer dynamics. These models frequently incorporate dynamic equations to simulate cancer cell populations at various differentiation stages and their interactions with the immune system \cite{reviewmath2}.

The overarching goal of mathematical cancer modelling is to understand and predict the behaviour of malignant cells and their interactions with the biological environment. However, constructing an effective model requires more than formulating equations; rigorous mathematical analysis is essential to ensure its robustness and reliability \cite{mathimp1,mathimp2,mathimp3}. This study examines critical aspects of a dynamic cancer model, including the existence and uniqueness of solutions, beside the  stability, feasibility, positivity and boundedness of the solutions.

Ensuring the uniqueness of solutions is vital, as it guarantees the model consistently produces identical outcomes under the same conditions, ensuring reproducibility and reliability. Stability plays an equally important role, as it ensures that small changes in initial conditions or parameters do not lead to drastically different outcomes. This is especially relevant in cancer systems, where precision in predictions is crucial despite variations in input data. Feasibility ensures that model solutions align with biological reality, accurately reflecting plausible cellular and tumour dynamics, thereby reinforcing the model's clinical relevance. Positivity is critical for maintaining biological validity, ensuring variables such as cell concentrations remain non-negative; otherwise, predictions would be meaningless in a biological context. Boundedness guarantees solutions remain within finite and realistic limits, preventing implausible or infinite outcomes that could compromise the model's credibility. Finally, the existence and uniqueness of solutions confirm the solvability of the model equations and ensure that they yield consistent results, which are indispensable for the model's utility in cancer research.

Together, these analyses establish a robust and reliable foundation for the proposed model, ensuring it is mathematically sound and practically relevant. This comprehensive approach provides a solid framework for advancing mathematical oncology and fostering future research in the field.

This article is organised as follows: Section $2$ introduces the mathematical model and the associated biological assumptions, and includes preliminary numerical simulations that motivate the subsequent theoretical analysis; Section $3$ establishes the feasibility of the model's solutions by proving positivity, boundedness, and the existence and uniqueness of solutions; in Section $4$, we perform a local stability analysis of the fixed points using Lyapunov exponents and Jacobian eigenvalues; finally, Section $5$ summarises the main findings and proposes directions for future research.

\section{The analyzed system\label{secoriginalmodel}}
In this work, we employ a model adapted from \cite{original}, which incorporates distinct populations of cancer cells (stem and differentiated) together with immune cell populations that target them. 
The cell populations are: Cancer stem cells ($S$); partial or totally differentiated cancer cells ($P$) (considered as  a single population); natural killers $(N\!K)$; suppressors derived from myeloid $(M)$; dendritic $(D)$; and cytotoxic $(T)$. To enhance the model's realism, we accounted for distinct cytotoxic cell populations targeting specific cancer cell types: $T\!S$ attacks stem-like cells $(S)$, while $T\!P$ targets differentiated cells $(P)$. Similarly, dendritic cells were modeled as subtypespecific effectors, with $D\!S$ acting against $S$ and $D\!P$ against $P$. The system of ordinary differential equations is shown in Equation \ref{sistema1}.

For a detailed explanation of the functions used in each equation, as well as their biological interpretation, please see \cite{original}. It should be noted that in \cite{originalesto} a first stability analysis was carried out from the stochastic point of view. However, in the present work it is a deterministic theoretical study, with complex mathematical tools.
\begin{equation}
\left\lbrace
\begin{array}{l}
\frac{d\,S}{d\,t}=\alpha_{s}\frac{S}{S+P} +\rho_{ps}P -(\rho_{sp}+\delta_{s})S - \beta_{s}T\!S\frac{S}{1+\frac{S^{\frac{1}{3}}}{l}} -\mu N\!K\frac{S}{1+\frac{S^{\frac{1}{3}}}{l}}\\\smallskip
\frac{d\,P}{d\,t}=  \alpha_{p}P(1-b P)  +(\alpha_{sp}  +2\rho_{sp})S- (\rho_{ps}+\delta_{p})P - \beta_{p}T\!P\frac{P}{1+\frac{P^{\frac{1}{3}}}{l}} -\mu N\!K\frac{P}{1+\frac{P^{\frac{1}{3}}}{l}}\\\smallskip
\frac{d\,T\!S}{d\,t}= k_{ts}\frac{D\!S}{D\!S +s_{ts}} - \delta_{ts}T\!S\\\smallskip
\frac{d\,T\!P}{d\,t}= k_{tp}\frac{D\!P}{D\!P +s_{tp}} - \delta_{tp}T\!P\\\smallskip
\frac{d\,D\!S}{d\,t}= \gamma_{ds}S- \beta_{ds}D\!ST\!S -\delta_{ds}D\!S\\\smallskip
 \frac{d\,D\!P}{d\,t}= \gamma_{dp}P- \beta_{dp}D\!PT\!P -\delta_{dp}D\!P\\\smallskip
\frac{d\,N\!K}{d\,t}= \sigma - fN\!K + g\frac{(S+P)^2}{(S+P)^2+h}-pN\!K\frac{S+P}{1+\frac{(S+P)^{\frac{1}{3}}}{l}}\\\smallskip
\frac{d\,M}{d\,t}= \rho_m - \beta_mM+ \alpha_m\frac{S+P}{S+P+q}
\end{array}
\right.
\label{sistema1}
\end{equation}

\begin{table}[width=.9\linewidth,cols=4,pos=h]
\caption{Parameters and their ranges.}\label{tabla1}
\begin{tabular*}{\tblwidth}{@{} CCC@{} }
\toprule
 Parameter & Description & Value \\
\midrule
$\alpha_s$ & Reproduction rate of S & $\left[0.14\, ,0.76\right] day^{-1}$ \\ 
$\alpha_{sp}$ & Production of $P$ through the asymmetrical division of $S$ & $\left[0.4\, ,0.76\right]day^{-1}$\\ 
$\alpha_p$ & Reproduction rate of $P$ & $\left[0\, ,0.8\right]day^{-1}$\\  
$b$ & Carrying capacity & $10^{-9}{\frac{1}{cells}}$\\  
$\rho_{ps}$ & Plasticity & $\left[0\, ,0.0564\right] day^{-1}$ \\ 
$\rho_{sp}$ & Production of $P$ through the total differentiation of $S$ & $\left[0\, ,0.76\right] day^{-1}$\\ 
$\delta_s$ & Death rate of $S$ due to natural processes & $\left[0\, ,0.25\right]day^{-1}$\\  
$\delta_p$ & Death rate of $P$ due to natural processes & $\left[0\, ,0.39\right]day^{-1}$\\ 
$l$ & Deep of the tumour accessible for immune cells & 100 $cell^{\frac{1}{3}}$\\ 
$\delta_{ds},\,\delta_{dp}$ & Death rate of $D\!S$ and $D\!P$ due to natural processes & $\left[0.2\, ,0.8\right]day^{-1}$\\  
$\beta_s,\, \beta_p$ & Death rate of cell type $S$ ($P$) due to $T\!S$ ($T\!P$)& $6.2\times 10^{-8}\, \frac{1}{T_i.day}$\\ 
$\mu$ & Fractional tumour cells kill rate by $N\!K$ cells & $3.23\times 10^{-7}\, \frac{cell}{day}$\\ 
$k_{ts},\,k_{tp}$  & Saturated rate of $T\!S$ ($T\!P$) due to activation by $D\!S$ ($D\!P$) &  $4.5\times 10^{4}\, \frac{T_i/ \mu L}{day}$\\ 
$s_{ts},\,s_{tp}$  & $D\!S$ ($D\!P$) EC50 for $T\!S$ ($T\!P$) activation rate &  $6.2\times 10^{-8}\, \frac{mDCs}{\mu L}$\\ 
$\delta_{ts},\,\delta_{tp}$ &  Death rate of $T\!S$ and $T\!P$ due to natural processes & 0.02 $day^{-1}$\\ 
$\gamma_{ds},\,\gamma_{dp}$ & Maturation rate of $D\!S$ ($D\!P$) by cancer cells consumption &$0.0063\, \frac{D_i / \mu L}{day/ (\mu L.day)}$\\ 
$\beta_{ds},\,\beta_{dp}$ & Death rate of cell type $D\!S$ ($D\!P$) due to $T\!S$ ($T\!P$) &$6.2\times 10^{-8}\, \frac{1}{T_i/ (\mu L.day)}$\\ 
$\sigma$ & Constant source of $N\!K$ cells & $1.4\times 10^{4}\frac{cell}{day}$\\ 
$f$ & Death rate of $N\!K$ cells & $4.12\times 10^{-2}day^{-1}$\\
$g$ & Maximum $N\!K$ cell recruitment rate & $2.5\times 10^{-2}day^{-1}$\\  
$h$ & Steepness coefficient of the $N\!K$ cell recruitment curve & $2.02\times 10^{7}cell^2$\\
$p$ & $N\!K$ cell inactivation rate by tumour cell & $1\times 10^{-7}\frac{1}{day.cell}$\\
$\rho_m$ & Normal $Ms$ production rate & $1.25\times 10^{6}\frac{cell}{day}$\\
$\beta_m$ & $Ms$ normal death rate &$0.25\frac{1}{day}$\\
$\alpha_m$ & $Ms$ expansion coefficient in tumour  &$1.2\times 10^{7}\frac{1}{cell.day}$\\
$q$ & Steepness coefficient of the $Ms$ production curve & $ 10^{10} cell$\\

\bottomrule
\end{tabular*}

\end{table}

\subsection{Numerical Simulations}
One way to explore the preliminary behaviour of the system is through numerical simulations, using the initial conditions considered in \cite{original}, which were selected for their biological plausibility and relevance to typical tumour-immune dynamics.These were performed considering  the values of the parameters in the Table \ref{tabla1} (for those parameters that are in a range, the average value was taken).

\begin{figure}
\centering
\includegraphics[scale=0.25]{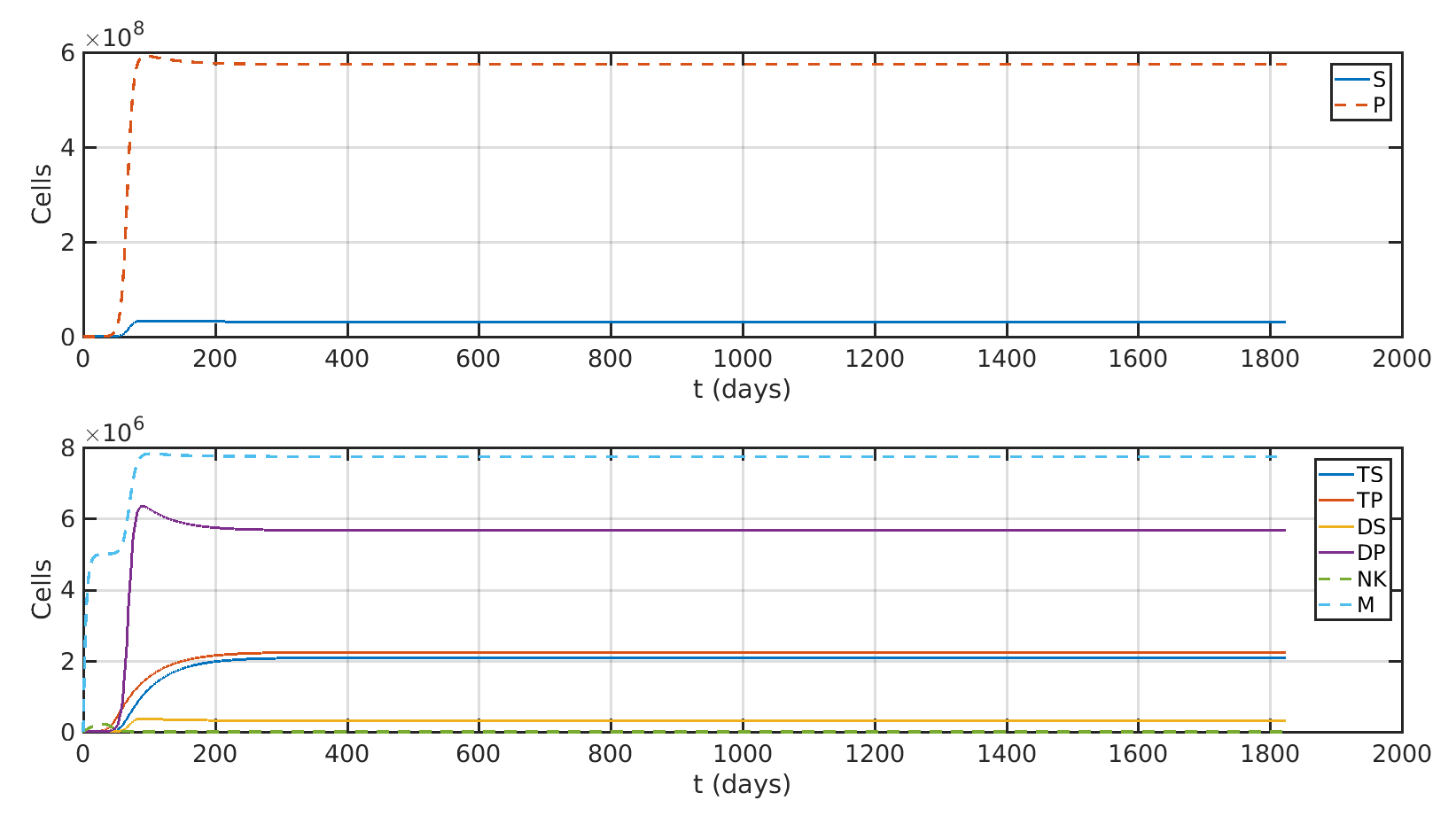}
\caption{(a) Time series for each type of cancer cells involved in the system. (b) Time series for each type of immune cells. Initial condition: $S(0)=100$, $P(0)=2000$, $T\!S(0)=T\!P(0)=D\!S(0)=D\!P(0)=N\!K(0)=M(0)=1000$.}\label{figu1}
\end{figure}

\begin{figure}
\centering
\includegraphics[scale=0.25]{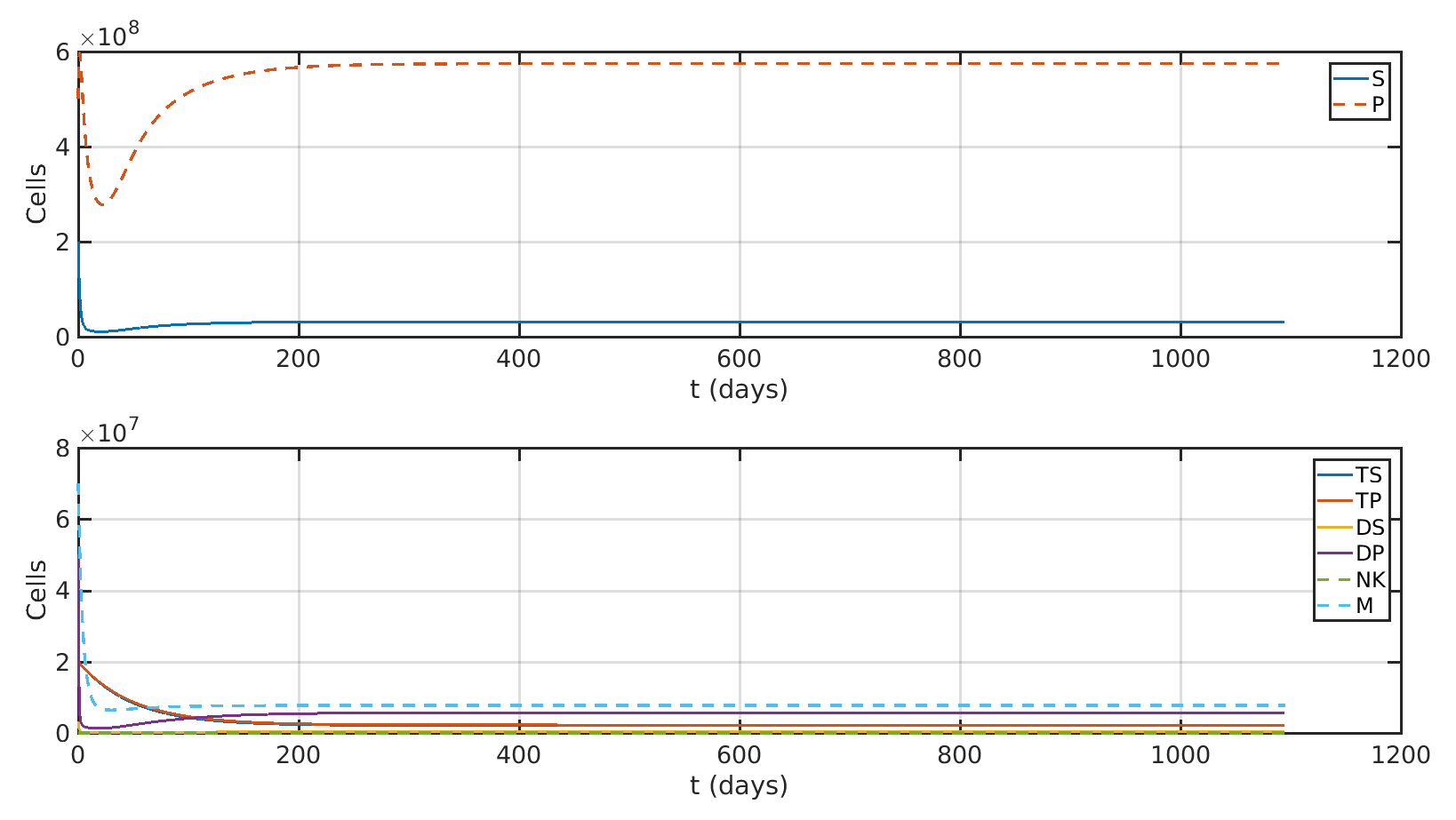}
\caption{(a) Time series for each type of cancer cells involved in the system. (b) Time series for each type of immune cells. Initial condition: $S(0)=2\times 10^8$, $P(0)=5\times 10^8$, $T\!S(0)=T\!P(0)=2\times 10^7$, $D\!S(0)= 3\times 10^6$, $D\!P(0)=5\times 10^7$, $N\!K(0)=2\times 10^4$ and $M(0)=7\times 10^7$.}\label{figu2}
\end{figure}

Here we will show two examples: for the Figure \ref{figu1}, the initial condition is $S(0)=100$, $P(0)=2000$ , $T\!S(0)=T\!P(0)=D\!S(0)=D\!P(0)=N\!K(0)=M(0)=1000$; for the Figure \ref{figu2}, the initial condition is $S(0)=2\times 10^8$, $P(0)=5\times 10^8$, $T\!S(0)=T\!P(0)=2\times 10^7$, $D\!S(0)= 3\times 10^6$, $D\!P(0)=5\times 10^7$, $N\!K(0)=2\times 10^4$ and $M(0)=7\times 10^7$.

Although numerical simulations suggest that the model tends to reach steady-state values, this visual observation is not sufficient to guarantee the system's stability or mathematical consistency. Due to the nonlinear nature and high dimensionality of the system, a rigorous mathematical validation is necessary. This motivates the subsequent analytical study of solution feasibility and stability provided in the following sections.

\subsection{Population $M$}
It should be noted that the first seven equations of the system \ref{sistema1} do not depend on the population $M$, so the feasibility analysis of the model can be performed without considering that population. Furthermore, it is important to note that it verifies the existence and uniqueness of solutions, beside the  stability, feasibility, positivity and boundedness.

Therefore, the system to be analyzed is
\begin{equation}
\left\lbrace
\begin{array}{l}
\frac{d\,S}{d\,t}=\alpha_{s}\frac{S}{S+P} +\rho_{ps}P -(\rho_{sp}+\delta_{s})S - \beta_{s}T\!S\frac{S}{1+\frac{S^{\frac{1}{3}}}{l}} -\mu N\!K\frac{S}{1+\frac{S^{\frac{1}{3}}}{l}}\\\smallskip
\frac{d\,P}{d\,t}=  \alpha_{p}P(1-b P)  +(\alpha_{sp}  +2\rho_{sp})S- (\rho_{ps}+\delta_{p})P - \beta_{p}T\!P\frac{P}{1+\frac{P^{\frac{1}{3}}}{l}} -\mu N\!K\frac{P}{1+\frac{P^{\frac{1}{3}}}{l}}\\\smallskip
\frac{d\,T\!S}{d\,t}= k_{ts}\frac{D\!S}{D\!S +s_{ts}} - \delta_{ts}T\!S\\\smallskip
\frac{d\,T\!P}{d\,t}= k_{tp}\frac{D\!P}{D\!P +s_{tp}} - \delta_{tp}T\!P\\\smallskip
\frac{d\,D\!S}{d\,t}= \gamma_{ds}S- \beta_{ds}D\!ST\!S -\delta_{ds}D\!S\\\smallskip
 \frac{d\,D\!P}{d\,t}= \gamma_{dp}P- \beta_{dp}D\!PT\!P -\delta_{dp}D\!P\\\smallskip
\frac{d\,N\!K}{d\,t}= \sigma - fN\!K + g\frac{(S+P)^2}{(S+P)^2+h}-pN\!K\frac{S+P}{1+\frac{(S+P)^{\frac{1}{3}}}{l}}
\end{array}
\right.
\label{sistema}
\end{equation}

\section{Model solution feasibility \label{maths}}

\subsection{Positivity of model's solutions}
Taking into account that the variables represent populations, all must be non-negative. In order to prove this, it is sufficient to demonstrate the following theorem:

\begin{thm} Let the initial conditions $S(0), P(0), T\!S(0), T\!P(0), D\!S(0), D\!P(0)$ and $N\!K(0)$ all be non - negative, then the solution set $\left(S(t), P(t), T\!S(t), T\!P(t), D\!S(t),\right.$ $\left. D\!P(t), N\!K(t)\right)$ of the System \ref{sistema} is positive $\forall t >0$.
\end{thm}\label{lema 1}

\begin{pf}
Let be the hyperparallelepiped $\Omega^{+} \subset \mathbb{R}^7_{\geq 0}$. The idea is to show that given an initial condition on the boundary of $\Omega^{+}$, the solutions remain in $\Omega^{+}$. 
The first equation of the System \ref{sistema} is
$$\frac{d\,S}{d\,t}=\alpha_{s}\frac{S}{S+P} +\rho_{ps}P -(\rho_{sp}+\delta_{s})S - \beta_{s}T\!S\frac{S}{1+\frac{S^{\frac{1}{3}}}{l}} -\mu N\!K\frac{S}{1+\frac{S^{\frac{1}{3}}}{l}}$$
Let us considered the hyperplan $S=0$ restricted to the domain $\Omega^{+}$. Then, we have 
\begin{equation}
\frac{d\,S}{d\,t}=\rho_{ps}P
\end{equation}
Since $\rho_{ps}>0$ y $P_0\leq 0$, $\frac{d\,S}{d\,t} \geq 0$. Therefore, $S$ is an increasing function. Therefore, $S(t)$ cannot be negative for any $t>0$.

Using analogous arguments, it can be shown that $P$, $T\!S$, $T\!P$, $D\!S$, $D\!P$ and $N\!K$ are increasing functions. In consequence, $P$, $T\!S$, $T\!P$, $D\!S$, $D\!P$ and $N\!K$  are non negative for any $t>0$.
\end{pf}

\subsection{Boundedness of model's solutions}

Here we show exists a positively invariant region with respect to the Model \ref{sistema}.
\begin{thm} There exists a bounded set $\Omega \subset \mathbb{R}^7_{\geq 0}$ - specifically, a hyperparallelepiped $H$ - where for all initial conditions $\left( S_0, P_0, T\!S_0, T\!P_0, D\!S_0, D\!P_0, N\!K_0\right) \in \Omega$ the solution to the Cauchy problem for the System \ref{sistema} remains bounded - with  $S_0, P_0$  nonzero simultaneously. 
\end{thm}\label{lem:3}

\begin{pf}
We begin by showing that for every initial condition in the positive hyperoctant $\Omega$, the solutions $T\!S(t),  T\!P(t)$ and $N\!K(t)$  are bounded for all $t\geq 0$.

\begin{itemize}

\item For the population $T\!S$, we have the differential equation:
$$\dot{T\!S}=k_{ts}\frac{D\!S}{D\!S +s_{ts}} - \delta_{ts}T\!S$$
Since $s_{ts}>0$ and from theorem \ref{lema 1}, it is given that $D\!S(t)\geq 0$, it follows that $\frac{D\!S}{D\!S +s_{ts}}<1$. Therefore,
$$\dot{T\!S}=k_{ts}\frac{D\!S}{D\!S +s_{ts}} - \delta_{ts}T\!S < k_{ts} - \delta_{ts}T\!S\Rightarrow \dot{T\!S} < k_{ts} - \delta_{ts}T\!S$$
This implies:
 \begin{equation}\label{if1}
	\dot{T\!S}+ \delta_{ts}T\!S< k_{ts} .
\end{equation}
Let the integrator factor be $exp\left(\int \delta_{ts}dt\right)$. Multiplying both sides of Inequation \ref{if1} by this factor, we obtain:
$$ exp\left(\int \delta_{ts}dt\right)\left(\dot{T\!S}+ \delta_{ts}T\!S\right)< k_{ts} exp\left(\int \delta_{ts}dt\right) \Leftrightarrow$$
\begin{equation}\label{if3}
	\frac{d}{dt}\left(T\!S(t)\cdot exp\left(\int \delta_{ts}dt\right)\right)< k_{ts} exp\left(\int \delta_{ts}dt\right).
\end{equation}
By integrating both sides in (\ref{if3}), 
\begin{equation}T\!S(t)\cdot exp\left( \delta_{ts}t\right)< \frac{k_{ts}}{\delta_{ts}} exp\left( \delta_{ts}t\right)+ C_1 \Leftrightarrow
0 <	T\!S(t)< \frac{k_{ts}}{\delta_{ts}} +const\cdot  \ exp\left( -\delta_{ts}t\right).
\end{equation}\label{if5}

As $T\!S(0)>0$, $0<T\!S(0)< \frac{k_{ts}}{\delta_{ts}} +C_1$. Which implies that constant is positive. Then, $T\!S(t) \leq \frac{k_{ts}}{\delta_{ts}} +C_1$.
Therefore, given any value $(S_0, P_0, T\!S_0; T\!P_0 D\!S_0, D\!P_0, N\!K_0) \in \mathbb{R}^7_{\geq 0}$, $T\!S(t)$ is upper bounded. This bound could determinete one of the faces of $\Omega$, which we denote as $T\!S^{\dagger} := \frac{k_{ts}}{\delta_{ts}} +C_1$. 
Moreover, $T\!S^{\dagger}$ could be as great as one would like, and this property will be used to determine the last two faces of $H$ with the $S^{\dagger}$ and $P^{\dagger}$ dimensions.

In an analogous way, given the value	$(S_0, P_0, T\!S_0; T\!P_0 D\!S_0, D\!P_0, N\!K_0) \in \mathbb{R}^7_{\geq 0}$, $T\!P(t)$ is upper bounded. This bound determines one of the faces of $\Omega$, which we denote as $T\!P^{\dagger}$.   

\item For the population $N\!K$, we have that
$$ \dot{N\!K}= \sigma - fN\!K + g\frac{(S+P)^2}{(S+P)^2+h}-pN\!K\frac{S+P}{1+\frac{(S+P)^{\frac{1}{3}}}{l}}$$
Since $h>0$ and from theorem \ref{lema 1}, it is given that $S(t),P(t)\geq 0$, it is follows that $\frac{(S+P)^2}{(S+P)^2+h}<1$, hence
$$\dot{N\!K}= \sigma - fN\!K + g\frac{(S+P)^2}{(S+P)^2+h}-pN\!K\frac{S+P}{1+\frac{(S+P)^{\frac{1}{3}}}{l}} < \sigma - fN\!K + g \Rightarrow \dot{N\!K}< \left(\sigma+g\right) - fN\!K$$
So, \begin{equation}\label{if1nk}
	\dot{N\!K}+ fN\!K< \left(\sigma+g\right) .
\end{equation}
Let the integrator factor be $exp\left(\int fdt\right)$ and multiplying both sides of Inequation \ref{if1nk} by this, we obtain:
\begin{equation}
	exp\left(\int fdt\right)\left(\dot{N\!K}+ fN\!K\right)<  \left(\sigma+g\right)  exp\left(\int fdt\right).
\end{equation}\label{if2nk}
Then,
\begin{equation}
	\frac{d}{dt}\left(N\!K(t)\cdot exp\left(\int fdt\right)\right)< \left(\sigma+g\right) exp\left(\int f dt\right).
\end{equation}\label{if3nk}
By integrating, 
\begin{equation}
	N\!K(t)\cdot exp\left( ft\right)< \frac{\sigma + g}{f} exp\left(f t\right)+ C_2.
\end{equation}\label{if4nk}
Then, 
\begin{equation}
	N\!K(t)< \frac{\sigma + g}{f} + C_2\cdot exp\left( -f t\right).
\end{equation}\label{if5nk}

As $N\!K(0)>0$, $0<N\!K(0)<  \frac{\sigma + g}{f}  +C_2$. Hence, $C_2>0$.
Consequently $N\!K(t) \leq  \frac{\sigma + g}{f}  + const$.  Therefore, given any value 	$(S_0, P_0, T\!S_0; T\!P_0 D\!S_0, D\!P_0, N\!K_0) \in \mathbb{R}^7_{\geq 0}$, $N\!K(t)$ is upper bounded. This bound determines one of the faces of $\Omega$, which we denote as $N\!K^{\dagger} := \frac{\sigma + g}{f} + C_2$.

\item To prove that the solutions $S(t)$ and 
$P(t)$ are bounded, we will do so indirectly.
To proceed with the demonstration, firstly, we fix one of the faces of the hyperparallelepiped H, for example, $S=S^{\dagger}$ where $S_0 < S^{\dagger}$. We have demonstrated $T\!S(t)$, $T\!P(t)$ and $N\!K(t)$ are bounded regardless of $S_0$ and $ P_0$. Consequently, among the variables that affect the search for bounded solutions to our Cauchy problem, we focus on the variable $P$.
The idea is to show that $S^{\dagger}$ determines a corresponding value $P = P^{\dagger}$, which will fix another face of the H and thus, determine the possible values $P_0$ such us $P_0 < P^{\dagger}$; in other words,  $P_0$ is a value to be determined within a range $[0; P^{\dagger}]$. 

Thus, it would be desirable that if the solution reaches that hyperplane $S=S^{\dagger}$, the derivative there is negative so that $S$ does not exceed value $S^{\dagger}$. Suppose that there is $t_1$ such as $S(t_1)= S^{\dagger}$ and it is intended to demonstrate that inequality holds for some value $P^{\dagger}$:
$$\dot{S}(t_1) = \alpha_{s} \frac{S^{\dagger}}{S^{\dagger}+P^{\dagger}} +\rho_{ps} P^{\dagger} - (\rho_{sp}+\delta_{s}) S^{\dagger} - \Big(\beta_{s} T\!S^{\dagger} 
+ \mu N\!K^{\dagger} \Big) \frac{S^{\dagger}}{1+\frac{\left(S^{\dagger}\right)^{\frac{1}{3}}}{l}}
\leq 0
$$
where $T\!S^{\dagger}$ and $N\!K^{\dagger}$ were previously described.

Given that $\frac{S^{\dagger}}{1+\frac{\left(S^{\dagger}\right)^{\frac{1}{3}}}{l}} \leq S^{\dagger}$, the prior expression is equivalent to:
\begin{equation}\label{if6}
\alpha_{s}\frac{S^{\dagger}}{S^{\dagger}+P^{\dagger}} \leq \underbrace{\Big( (\rho_{sp}+\delta_{s}) + \beta_{s}T\!S^{\dagger} + \mu N\!K^{\dagger} \Big) S^{\dagger} }_{Const.} - \rho_{ps} P^{\dagger} 
\end{equation}

The homographic function $f\left(P^{\dagger}\right)=	\alpha_{s} \frac{S^{\dagger}}{S^{\dagger} + P^{\dagger}}$ decreases asymptotically to $0$ with $S^{\dagger}$ fixed, while the right-hand side is a line with slope $-\rho{ps}$.
Since $-\rho{ps} < 0$ there is a single positive intersection point $P^{\ddagger}$.
However, this inequality does not hold for all $P^{\dagger} \in \mathbb{R}_{\geq 0}$. For $P^{\dagger} > P^{\ddagger}$, the inequality reverse ($\dot{S}(t_1) > 0$), which would allows $S(t)$ to grow beyond $S^{\dagger}$.
Without loss of generality, the Figure \ref{cota} illustrates the general scenario described by Inequality \ref{if6}: 
\begin{figure}[h]
	\centering
	\includegraphics[scale=0.28]{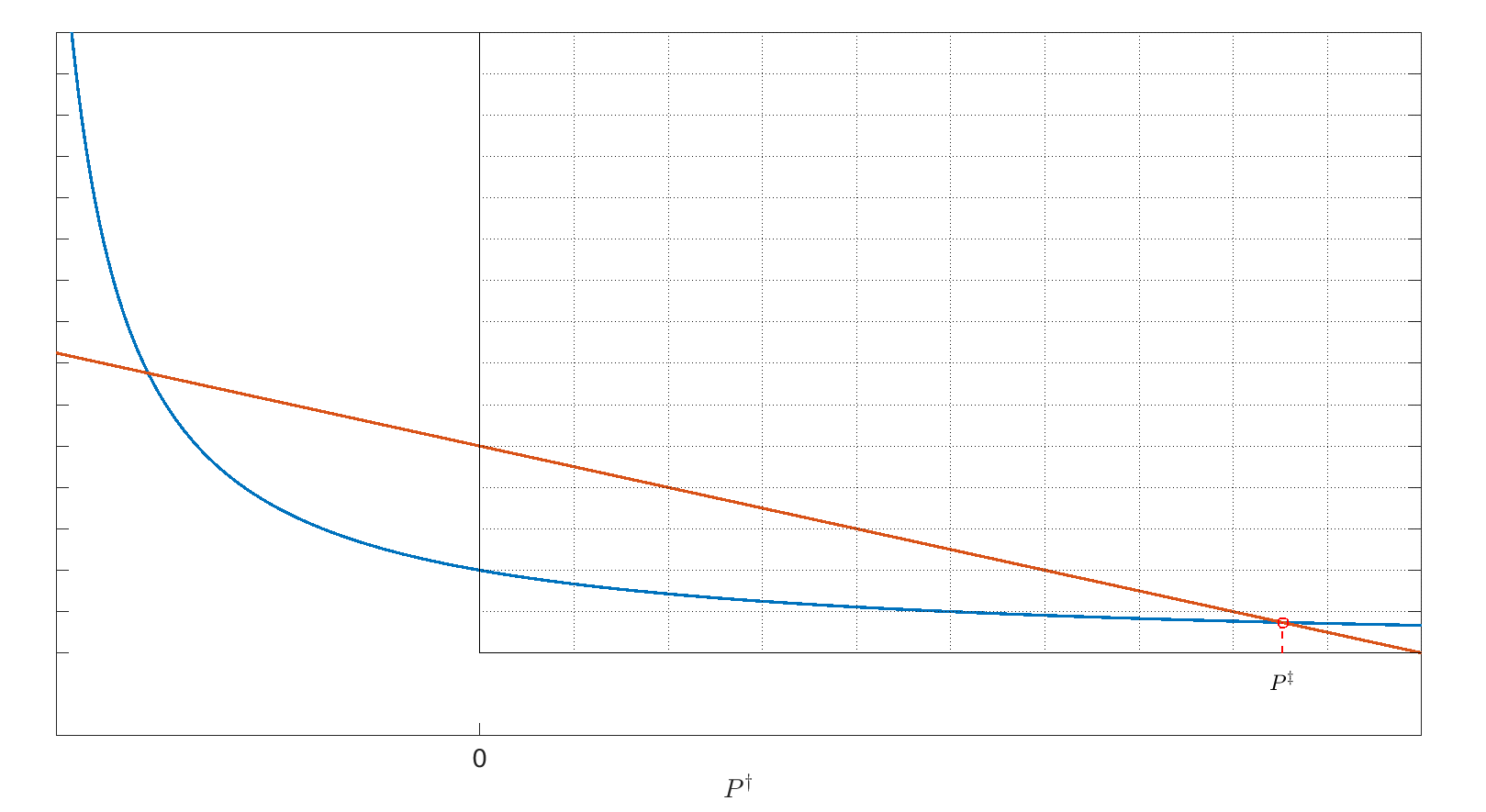}    
\caption{The abscissa axis represents the solution to Inequality \ref{if6}, where the only unknown is $P^{\dagger}$. The intersection point between the homographic function and the linear function determines $P^{\ddagger}$. The dashed region is the one that makes biological sense for the problem.}\label{cota}
\end{figure}

Therefore, given $S^{\dagger}$ it is possible to find $P^{\dagger} < P^{\ddagger}$, such as $\dot{S}(t_1) \leq 0$ is true. So if the solution reaches the plane $S=S^{\dagger}$, it does not exceed it. Since $T\!S_0$ and $N\!K_0$ can be arbitrarily large, the value $P^{\ddagger}$ (the intersection point) can grow without bound.  This observation will be revisit in the following item.

\item For $P$, we start with the differential equation:
$$\dot{P}(t)=  \alpha_{p}P(1-b P)  +(\alpha_{sp}  +2\rho_{sp})S- (\rho_{ps}+\delta_{p})P -\frac{P}{1+\frac{P^{\frac{1}{3}}}{l}}\left( \beta_{p}T\!P +\mu N\!K\right)$$
Where the term $\alpha_{p}P(1-b P)$  is maximized when $P = \frac{\alpha_{p}}{4b}$, yielding: 
$$\alpha_{p}P(1-b P) \leq \frac{\alpha_{p}}{4b}$$
Since $S(t) < S^{\dagger}$ (as previously established for $S(t))$, we substitute $S$ with its upper bound $S^{\dagger}$ and drop the negative term involving $T\!P$ and $N\!K$  (as it only decreases $\dot{P}$):
\begin{equation}
	\dot{P}\leq \frac{\alpha_{p}}{4b} +(\alpha_{sp}  +2\rho_{sp})S^{\dagger}- (\rho_{ps}+\delta_{p})P.
\end{equation}\label{ifpp}
This simplifies to the inequality:
\begin{equation}\label{pp2}
	\dot{P} + (\rho_{ps} + \delta_{p}) P \leq \frac{\alpha_{p}}{4b} + (\alpha_{sp} + 2\rho_{sp}) S^{\dagger}.
\end{equation}
Multiplying both sides of Inequation \ref{pp2} by the integrator factor $exp\left(\int (\rho_{ps}+\delta_{p})dt\right)$:

\begin{equation}
	\frac{d}{dt}\left(P(t)\cdot exp\left(\int (\rho_{ps}+\delta_{p})dt\right)\right) \leq \left(\frac{\alpha_{p}}{4b} +(\alpha_{sp}  +2\rho_{sp})S^{\dagger}\right)exp\left(\int (\rho_{ps}+\delta_{p})dt\right).
\end{equation}\label{pp4}
Integrating both sides from $0$ to $t$:
\begin{equation}
	P(t)\cdot exp\left((\rho_{ps}+\delta_{p})t\right) \leq \frac{\alpha_{p}+4b(\alpha_{sp}  +2\rho_{sp})S^*}{4b(\rho_{ps}+\delta_{p})} exp\left((\rho_{ps}+\delta_{p})t\right) + C_3.
\end{equation}\label{pp5}
where $C_4$ is a constant determined by initial conditions.
Divide through by $exp\left((\rho_{ps}+\delta_{p})t\right)$:
\begin{equation}\label{pp6}
	P(t)\leq \frac{\alpha_{p}+4b(\alpha_{sp}  +2\rho_{sp})S^{\dagger}}{4b(\rho_{ps}+\delta_{p})}  + C_3 \ exp\left(-(\rho_{ps}+\delta_{p})t\right).
\end{equation}
As $P(0)>0$, $0<P(0)\leq \frac{\alpha_{p}+4b(\alpha_{sp}  +2\rho_{sp})S^{\dagger}}{4b(\rho_{ps}+\delta_{p})}  + C_3$. Which implies that $C_3>0$.

Then $P(t)$ verifies:
$$P(t) \leq \frac{\alpha_{p}+4b(\alpha_{sp}  +2\rho_{sp})S^{\dagger}}{4b(\rho_{ps}+\delta_{p})}+C_3.$$
Thus, $P(t)$ is uniformly bounded for all $t \geq 0$, provided by if:
$$\frac{\alpha_{p}+4b(\alpha_{sp}  +2\rho_{sp})S^{\dagger}}{4b(\rho_{ps}+\delta_{p})} +C_3 < P^{\ddagger}$$

However, this limitation is valid, as noted in the previous point, $P^{\ddagger}$ can be arbitrarily large depending on the values $T\!P^{\dagger}$ and $N\!K^{\dagger}$. 
The model shows that the existence of $H$ is consistent with what has been observed experimentally:  there are several works that propose a mathematical model of tumor growth, compare it with experimental data and observe that it fits well. After some time $t$, these models present a plateau. Some examples are the work done in \cite{SPacotada, SPacotada1, SPacotada2}.


\item For the population $D\!S$, we have that
$$\dot{D\!S}= \gamma_{ds}S- \beta_{ds}D\!S T\!S -\delta_{ds}D\!S$$

Let $S^{\dagger}$ be the upper limit of $S$. As $T\!S\leq \frac{k_{ts}}{\delta_{ts}} + const = T\!S^{\dagger}$ thus, 
$$\dot{D\!S}= \gamma_{ds}S- \beta_{ds}D\!S T\!S -\delta_{ds}D\!S < \gamma_{ds}S^{\dagger} -\delta_{ds}D\!S $$
By rearranging, \begin{equation}\label{if6ds}
\dot{D\!S} +\delta_{ds}D\!S < \gamma_{ds}S^{\dagger}.
\end{equation}

Be the integral factor $exp\left(\int \delta_{ds}dt\right)$ and multiplying by this on both sides of Inequation \ref{if6ds}, we obtain
\begin{equation}
exp\left(\int \delta_{ds}dt\right)\left(\dot{D\!S} +\delta_{ds}D\!S\right)< \gamma_{ds}S^{\dagger}exp\left(\int \delta_{ds}dt\right).
\end{equation}\label{if7}
Then, 
\begin{equation}
\frac{d}{dt}\left(D\!S(t)\cdot exp\left(\int \delta_{ds}dt\right)\right)< \gamma_{ds}S^{\dagger}exp\left(\int \delta_{ds}dt\right).
\end{equation}\label{if8}
By integrating from $0$ to $t$, 
\begin{equation}
D\!S(t)\cdot exp\left(\delta_{ds}t\right)< \frac{\gamma_{ds}S^{\dagger}}{\delta_{ds}}exp\left(\delta_{ds}t\right) +C_4.
\end{equation}\label{if9}
Dividing by $exp\left(\delta_{ds}t\right)$
\begin{equation}
D\!S(t)< \frac{\gamma_{ds}S^{\dagger}}{\delta_{ds}} +C_4\cdot exp\left(-\delta_{ds}t\right).
\end{equation}\label{if10}
Then $D\!S(t)< \frac{\gamma_{ds}S^{\dagger}}{\delta_{ds}} +C_4$.

As $D\!S(0)>0$, $0<D\!S(0)<  \frac{\gamma_{ds}S^{\dagger}}{\delta_{ds}} +C_4$. Hence, $C_4>0$. Then, given the value $(S_0, P_0, T\!S_0; T\!P_0 D\!S_0, D\!P_0, N\!K_0) \in \mathbb{R}^7_{\geq 0}$, $D\!S(t)$ is upper bounded. Similarly, $D\!P(t)$ is upper bounded.
\end{itemize}

Finally, all the solutions of the model are bounded.
\end{pf}

\subsection{Existence and uniqueness of model's solution}\label{exisuni}

The Picard-Linderl\"{o}f theorem established the existence and uniqueness of solution:
\begin{thm}\label{theolip}
Let $D\subseteq \mathbb{R}\times \mathbb{R}^n$ be an open set $f:D\to\mathbb{R}$ be a continuous function. Consider the initial value problem (IVP):

\begin{equation}
\frac{dx}{dt}=f(t,x) \qquad x\left(t_0\right)=x_0\label{ma1}
\end{equation}
where $\left(t_0,x_0\right)\in D$.

if $f$ is locally Lipschitz continuous in $x$, i.e., for every compact subset $K \subset D$, there exists a constant $L>0$ such that 
\begin{equation}\label{lips}
\Vert f\left(t,x_1\right)-f\left(t,x_2\right)\Vert \leq L\Vert x_2 - x_2\Vert \qquad \forall \left(t,x_1\right),\left(t,x_2\right)\in K
\end{equation}
then:
\begin{itemize}
\item Existence: there exists an interval $I=\left[t_0-\delta, t_0+\delta\right]$, for some $\delta>0$, on which a solution $x(t)$ to the IVP exists.
\item Uniceness: the solution is unique on $I$, meaning that if $x_1(t)$ y $x_2(t)$ both solve the IVP on $I$, then $x_1(t)=x_2(t)$ for all $t\in I$.
\end{itemize}
\end{thm}

It is important to remember that Inequality \ref{lips} is achieved if $\frac{\partial f}{\partial x_i}$ for $i=1,\ldots,n$ are continuous and bounded functions in $K$.

Let $G = \left(g_1,\ldots,g_7\right)$ where $G$ is the velocity field defined  by the System \ref{sistema}. Then 

\begin{itemize}
\item For $g_1$:
\begin{equation}
\left\lbrace
\begin{array}{ll}
\frac{\partial g_1}{\partial S} =\frac{\alpha_s P}{(P+S)^2}-\rho_{sp}-\delta_s-\left(\beta_s T\!S+\mu N\!K\right) \frac{1+\frac{2}{3}\frac{S^{\frac{1}{3}}}{l}}{\left(1+\frac{S^{\frac{1}{3}}}{l}\right)^2}, &  \left|\frac{\partial g_1}{\partial S}\right| < \infty\\\smallskip 
\frac{\partial g_1}{\partial P} = \rho_{ps}-\frac{\alpha_s S}{(P+S)^2}, & \left|\frac{\partial g_1}{\partial P} \right|= \left|\rho_{ps}-\frac{\alpha_s S}{(P+S)^2}\right|<\infty \\\smallskip
\frac{\partial g_1}{\partial T\!S} = - \beta_s\frac{S}{1+\frac{S^{\frac{1}{3}}}{l}}, &\left|\frac{\partial g_1}{\partial T\!S}\right|= \left|- \beta_s\frac{S}{1+\frac{S^{\frac{1}{3}}}{l}} \right| < \infty\\\smallskip
\frac{\partial g_1}{\partial N\!K} = -\mu\frac{S}{1+\frac{S^{\frac{1}{3}}}{l}}, &  \left|\frac{\partial f_1}{\partial D}\right|= \left|-\mu\frac{S}{1+\frac{S^{\frac{1}{3}}}{l}} \right| < \infty\\
\end{array}
\right.\label{f1}
\end{equation}
and $\frac{\partial g_1}{\partial T\!P}=0$, $\frac{\partial g_1}{\partial D\!S}=0$ and $\frac{\partial g_1}{\partial D\!P}=0$.

\smallskip

\item For $g_2$:
\begin{equation}
\left\lbrace
\begin{array}{ll}
\frac{\partial g_2}{\partial S} =\alpha_{sp}  +2\rho_{sp},& \left|\frac{\partial g_2}{\partial S}\right|= \left|\alpha_{sp}  +2\rho_{sp} \right| < \infty \\\smallskip 
\frac{\partial g_2}{\partial P} = \alpha_p\left(1-2bP\right)-\rho_{ps}-\delta_P-\left(\beta_p  T\!P+\mu N\!K\right) \frac{1+\frac{2}{3}\frac{P^{\frac{1}{3}}}{l}}{\left(1+\frac{P^{\frac{1}{3}}}{l}\right)^2},&  \left|\frac{\partial g_2}{\partial P}\right|< \infty\\\smallskip
\frac{\partial g_2}{\partial T\!P} = -\beta_p\frac{P}{1+\frac{P^{\frac{1}{3}}}{l}}, & \left|\frac{\partial g_2}{\partial T\!P}\right|= \left|-\beta_p\frac{P}{1+\frac{P^{\frac{1}{3}}}{l}} \right| < \infty \\\smallskip
\frac{\partial g_2}{\partial N\!K} = -\mu\frac{P}{1+\frac{P^{\frac{1}{3}}}{l}}, &  \left|\frac{\partial g_2}{\partial N\!K}\right|= \left|-\mu\frac{P}{1+\frac{P^{\frac{1}{3}}}{l}} \right| < \infty\\
\end{array}
\right.\label{f2}
\end{equation}
and $\frac{\partial g_2}{\partial T\!S}=0$, $\frac{\partial g_2}{\partial D\!S}=0$ and $\frac{\partial g_2}{\partial D\!P}=0$.

\smallskip

\item For $g_3$:
\begin{equation}
\left\lbrace
\begin{array}{ll}
\frac{\partial g_3}{\partial T\!S} =-\delta_{ts}, & \left|\frac{\partial g_3}{\partial T\!S}\right|= \left|-\delta_{ts} \right| < \infty \\\smallskip 
\frac{\partial g_3}{\partial D\!S} = k_{ts}\frac{s_{ts}}{\left(D\!S+s_{ts}\right)^2}, &  \left|\frac{\partial g_3}{\partial D\!S}\right|= \left|k_{ts}\frac{s_{ts}}{\left(D\!S+s_{ts}\right)^2} \right| < \infty \\
\end{array}
\right.\label{g3}
\end{equation}
and $\frac{\partial g_3}{\partial S}=0$, $\frac{\partial g_3}{\partial P}=0$, $\frac{\partial g_3}{\partial T\!P}=0$, $\frac{\partial g_3}{\partial D\!P}=0$ and  $\frac{\partial g_3}{\partial N\!K}=0$.

\smallskip

\item For $g_4$:
\begin{equation}
\left\lbrace
\begin{array}{ll}
\frac{\partial g_4}{\partial T\!P} =-\delta_{tp}, & \left|\frac{\partial g_4}{\partial T\!S}\right|= \left|-\delta_{tp} \right| < \infty \\\smallskip 
\frac{\partial g_4}{\partial D\!P} = k_{tp}\frac{s_{tp}}{\left(D\!P+s_{tp}\right)^2}, &  \left|\frac{\partial g_4}{\partial D\!P}\right|= \left|k_{tp}\frac{s_{tp}}{\left(D\!P+s_{tp}\right)^2} \right| < \infty \\
\end{array}
\right.\label{g4}
\end{equation}
and $\frac{\partial g_4}{\partial S}=0$, $\frac{\partial g_4}{\partial P}=0$, $\frac{\partial g_4}{\partial T\!S}=0$, $\frac{\partial g_4}{\partial D\!S}=0$ and $\frac{\partial g_4}{\partial N\!K}=0$.

\smallskip

\item For $g_5$:
\begin{equation}
\left\lbrace
\begin{array}{ll}
\frac{\partial g_5}{\partial S} = \gamma_{ds}, & \left|\frac{\partial g_5}{\partial S}\right|= \left|\gamma_{ds} \right| < \infty \\\smallskip 
\frac{\partial g_5}{\partial T\!S} = -\beta_{ds}D\!S, &  \left|\frac{\partial g_5}{\partial T\!S}\right|= \left|-\beta_{ds}D\!S \right| < \infty \\\smallskip
\frac{\partial g_5}{\partial D\!S} = -\beta_{ds}T\!S-\delta_{ds}, &  \left|\frac{\partial g_5}{\partial D\!S}\right|= \left| -\beta_{ds}T\!S-\delta_{ds}\right| < \infty\\
\end{array}
\right.\label{g5}
\end{equation}

and $\frac{\partial g_5}{\partial P} =0$, $\frac{\partial g_5}{\partial T\!P} =0$,  $\frac{\partial g_5}{\partial D\!P} =0$ and $\frac{\partial g_5}{\partial N\!K} =0$.

\smallskip

\item For $g_6$:
\begin{equation}
\left\lbrace
\begin{array}{ll}
\frac{\partial g_6}{\partial P} = \gamma_{dp}, & \left|\frac{\partial g_6}{\partial P}\right|= \left|\gamma_{dp} \right| < \infty \\\smallskip 
\frac{\partial g_6}{\partial T\!P} = -\beta_{dp}D\!P, &  \left|\frac{\partial g_6}{\partial T\!P}\right|= \left|-\beta_{dp}D\!P \right| < \infty \\\smallskip
\frac{\partial g_6}{\partial D\!P} = -\beta_{dp}T\!P-\delta_{dp}, &  \left|\frac{\partial g_6}{\partial D\!P}\right|= \left|-\beta_{dp}T\!P-\delta_{dp}\right| < \infty\\
\end{array}
\right.\label{g6}
\end{equation}

and $\frac{\partial g_6}{\partial S} =0$, $\frac{\partial g_6}{\partial T\!S} =0$,  $\frac{\partial g_6}{\partial D\!S} =0$ and $\frac{\partial g_6}{\partial N\!K} =0$.
\smallskip

\item For $g_7$:
\begin{equation}
\left\lbrace
\begin{array}{ll}
\frac{\partial g_7}{\partial S} = g\frac{2(S+P)(h-1)}{\left((S+P)^2+h\right)^2}-\mu N\!K\frac{1+\frac{2}{3}\frac{(S+P)^{\frac{1}{3}}}{l}}{\left(1+\frac{(S+P)^{\frac{1}{3}}}{l}\right)^2}, & \left|\frac{\partial g_7}{\partial S}\right|= \left| g\frac{2(S+P)(h-1)}{\left((S+P)^2+h\right)^2}-\mu N\!K\frac{1+\frac{2}{3}\frac{(S+P)^{\frac{1}{3}}}{l}}{\left(1+\frac{(S+P)^{\frac{1}{3}}}{l}\right)^2} \right| < \infty \\\smallskip 
\frac{\partial g_7}{\partial P} = g\frac{2(S+P)(h-1)}{\left((S+P)^2+h\right)^2}-\mu N\!K\frac{1+\frac{2}{3}\frac{(S+P)^{\frac{1}{3}}}{l}}{\left(1+\frac{(S+P)^{\frac{1}{3}}}{l}\right)^2}, &  \left|\frac{\partial g_7}{\partial P}\right|= \left|g\frac{2(S+P)(h-1)}{\left((S+P)^2+h\right)^2}-\mu N\!K\frac{1+\frac{2}{3}\frac{(S+P)^{\frac{1}{3}}}{l}}{\left(1+\frac{(S+P)^{\frac{1}{3}}}{l}\right)^2} \right| < \infty \\\smallskip
\frac{\partial g_7}{\partial N\!K} = -f-p\frac{S+P}{1+\frac{(S+P)^{\frac{1}{3}}}{l}}, &  \left|\frac{\partial g_7}{\partial N\!K}\right|= \left|-f-p\frac{S+P}{1+\frac{(S+P)^{\frac{1}{3}}}{l}}\right| < \infty\\
\end{array}
\right.\label{g7}
\end{equation}
and $\frac{\partial g_7}{\partial T\!S}=0$, $\frac{\partial g_7}{\partial T\!P}=0$, $\frac{\partial g_7}{\partial D\!S}=0$ and $\frac{\partial g_7}{\partial D\!P}=0$.

\end{itemize}

Finally, we  proved that each partial derivatives is continuous and bounded. Therefore 
we can conclude from the Theorem \ref{theolip} that there exist only one solution to the System \ref{sistema} in the domain $K$.

\section{Local stability analysis of the fixed points\label{local}}

The stable fixed points of our autonomous system correspond to the asymptotic behavior of numerical solutions in a neighborhood of constant solutions. In rare cases, these points can be expressed using explicit analytical formulas based on the model parameters, allowing for exact characterization. However, in most nonlinear systems, such as ours, such accuracy is unattainable. This impossibility manifests itself in our system in two ways, as we will explain below.
\begin{itemize}
\item Intrinsic complexity of the system: The high nonlinearity and the number of variables make exact convergence to the fixed point difficult.

\item Numeric machine precision: When working with double-precision arithmetic, the representation limit is on the order of $15-17$ significant decimal digits. Therefore, the numerical solutions obtained are conditioned by this barrier inherent to the floating-point format.
\end{itemize}
With these clarifications, our fixed point stabilized at the values shown in Table \ref{fprs}.

\begin{table}[width=.9\linewidth,cols=3,pos=h]
\caption{Fixed points for the model.}\label{fprs}
\begin{tabular*}{\tblwidth}{@{} CCC@{} }
\toprule
 Population & Fixed point symbol & Value (cells) \\
 \midrule
$S$ & $S^*$  & $3.051897478716737\times 10^7$ \\ 
$P$ & $P^*$ & $5.806341558347363\times 10^8$\\ 
$T\!S$ & $T\!S^*$ & $2.079904846837571\times 10^6$\\  
$T\!P$ & $T\!P^*$ & $2.240218269209419\times 10^6$\\  
$D\!S$ & $D\!S^*$ & $3.056972535914865\times 10^5$\\ 
$D\!P$ & $D\!P^*$ & $5.725516059540588\times 10^6$\\  
$N\!K$ & $N\!K^*$ & $2.159263390129249\times 10^3$\\ 
\bottomrule
\end{tabular*}
\end{table}

As for the stability of this fixed point, due to the limitations mentioned above, we inevitably resort to numerical methods.

One way to analyse the stability of the fixed points is by calculating the Largest Lyapunov Exponent $\lambda_i$ \cite{lyapunov,lyapunov2}. This a measure that provides a notion of the divergence or convergence of nearby trajectories for the dynamical system. The Largest Lyapunov Exponent for a system is approximated by:$$ \delta d(t) \approx exp(\lambda_i t) \delta d_0$$
where  $\delta d_0$ is the initial separation of the nearest neighbors and $\delta d(t)$  is their distance after time evolution $t$.  If $\lambda_i < 0$ for all system components, the system is asymptotically stable. A positive $\lambda_i$ indicates chaos or instability. Table \ref{lle} shows the Largest Lyapunov Exponents for all populations in this model, therefore the $\left(S^*, P^*,T\!S^*,T\!P^*,D\!S^*, D\!P^*, N\!K^*\right)$ is asymptotic stable.

\begin{table}[width=.9\linewidth,cols=2,pos=h]
\caption{Largest Lyapunov Exponents.}\label{lle}
\begin{tabular*}{\tblwidth}{@{} CC@{} }
\toprule
 Population & $\lambda$ \\
 \midrule 
$S$   & $-0.000169874$ \\ 
$P$  & $-0.000168010$\\ 
$T\!S$  & $-0.000154113$\\  
$T\!P$  & $-0.000435045$\\  
$D\!S$  & $-0.000245745$ \\ 
$D\!P$  & $-0.000214436$\\ 
$N\!K$  & $-0.067451975$\\  
\bottomrule
\end{tabular*}
\end{table}

Another way to analyze the stability is by finding the eigenvalues of the associated Jacobian matrix evaluated at the found fixed point. The Jacobian matrix is given by the partial derivative of $G$, where $G = \left(g_1,\ldots,g_7\right)$ as defined in Subsection \ref{exisuni}. The eigenvalues at $\left(S^*, P^*,TS^*, T\!P^*, D\!S^*, D\!P^*, N\!K^*\right)$ are $\lambda_1 \approx -0.526430$, $\lambda_2 \approx -0.309199$, $\lambda_3 \approx -0.019994$, $\lambda_4 \approx -0.020002$, $\lambda_5 \approx -0.628963 $, $\lambda_6 \approx -0.638887$ and $\lambda_7 \approx -6.483704$ . Since all the eigenvalues are negative, the fixed point is locally asymptotically stable.

Therefore, by using two different tools, it can be ensured that the fixed point found is stable.

\section{Conclusions}

The analysis of a model simulating the interactions between cell populations not only contributes to the field of mathematical oncology but also reinforces the field of applied mathematics. Conducting a rigorous functional analysis of the model is crucial, as it strengthens its mathematical validity and enhances its effectiveness and predictability within the biological context. Specifically, this work examines the feasibility and stability of a complex integrative model of cancer cell differentiation. This model incorporates populations of cancer cells at various stages of differentiation, as well as immune cells, to describe the dynamic interactions between cancerous cells and their surrounding environment.
Additionally, we provide a rigorous proof of the model's consistency, demonstrating that the proposed framework is mathematically sound and capable of yielding reliable results. This validation ensures that the model accurately reflects the biological processes it aims to simulate, making it a robust tool for further research and applications in mathematical oncology.
An interesting proposal for future work is to conduct an analysis similar to the one presented in this study but considering a broader range of interactions between different cell populations. A key focus would be to examine the potential stability and/or shifts of the fixed points, along with understanding their corresponding biological significance. This analysis could provide deeper insights into how changes in the interactions might affect the overall behaviour of the system, including its response to fluctuations and perturbations, thus offering a more comprehensive view of cancer-immune system interactions.

\section*{Acknowledgement}

This work is financed by the Multiannual Research Projects (PIP) N$^{\circ}$ 11220200100439CO of the Consejo Nacional de Investigaciones Cient\'{i}ficas y T\'{e}cnicas (CONICET), Argentina .

Funded by the European Union. Views and opinions expressed are however those of the author(s) only and do not necessarily reflect those of the European Union or European Research Executive Agency (REA) (granting authority). Neither the European Union nor the granting authority can be held responsible for them. The project leading to this application has received funding from the European Union's Horizon Europe programme under the MSCA-SE grant agreement N$^{\circ}$ 101131463 - SIMBAD: Statistical Inference from Multiscale Biological Data: Theory, Algorithms, Applications.

\printcredits

\bibliographystyle{model1-num-names}

\bibliography{Article}

\end{document}